%% file: Paper_Oldemar_Rodriguez.tex
\documentclass[graybox]{svmult}

\usepackage{amsfonts}
\usepackage{mathptmx}       
\usepackage{helvet}         
\usepackage{courier}        
\usepackage{type1cm}        
\usepackage{makeidx}         
\usepackage{graphicx}        
\usepackage{multicol}        
\usepackage[bottom]{footmisc}

\makeindex             

\begin{document}

\title*{Riemannian Statistics for Any Type of Data}

\author{Oldemar Rodríguez Rojas}

\institute{Oldemar Rodríguez Rojas \at School of Mathematics, CIMPA, University of Costa Rica, San José, Costa Rica, \email{oldemar.rodriguez@ucr.ac.cr}}

\maketitle

\abstract{This paper introduces a novel approach to statistics and data analysis, departing from the conventional assumption of data residing in Euclidean space to consider a Riemannian Manifold. The challenge lies in the absence of vector space operations on such manifolds. Pennec X. et al. in their book Riemannian Geometric Statistics in Medical Image Analysis proposed analyzing data on Riemannian manifolds through geometry, this approach is effective with structured data like medical images, where the intrinsic manifold structure is apparent. Yet, its applicability to general data lacking implicit local distance notions is limited. We propose a solution to generalize Riemannian statistics for any type of data.} 


\section{Introduction}
\label{sec:int}

Each point in a data table can be imagined as a star or planet in the universe, especially when dealing with big data issues. In the universe, due to the infinitely different sizes of constellations, there are vastly different perceptions of distances between celestial bodies. For example, two constellations or galaxies that appear to be the same size from a distance (from Earth, for example) could be infinitely different, and one could even fit inside the other in a very small portion or empty space within it. Similarly, in data, there are local notions of distance corresponding to different regions of the data, and this should be considered when calculating statistical indices or models. For this reason, especially in problems involving Big Data, {\em thinking that the data is in Euclidean space is just as wrong as thinking that the earth is flat}. To address this, we propose considering that the data exists within a Riemannian manifold, where these local notions of distance can be effectively taken into account. 

In  \cite{pennec} {\em Pennec et al.}, the authors had proposed the idea of analyzing data on Riemannian manifolds through the use of geometry. However, this idea is not readily applicable to general data where there are no implicit notions of local distance. The idea that we propose go beyond, our main idea is to endow a Riemannian manifold structure to any given set of data. This approach leads to a significant improvement in the results of various statistical analyses as well as their interpretability.

In another significant contribution, McInnes et al. \cite{mcinnes} introduce UMAP (Uniform Manifold Approximation and Projection), a novel technique for manifold learning and dimension reduction. Utilizing simplicial complexes, Čech complexes, and the Nerve theorem, UMAP gains additional benefits from this Riemannian metric-based approach. It generates a local metric space associated with each point, allowing for meaningful distance measurements. Consequently, the algorithm can assign weights to edges in a graph (simplicial complex), signifying the local metric-based separation between the original points. So the idea that we proposed in this paper is to use the local notions of distance that the UMAP algorithm generates in any data table to provide the it with local distances. 

UMAP, as a successor to $t$-SNE method, inherits a controversy associated with the $t$-SNE method.  A comprehensive discussion on this topic can be found at \url{https://umap-learn.readthedocs.io/en/latest/clustering.html}.
Despite these concerns, there are still valid reasons to utilize UMAP as a preprocessing step for clustering. As highlighted in the discussion, when applied to real high-dimensional datasets such as {\tt MNIST} data \cite{deng} or {\tt cell RNA-seq} data \cite{karthik}, and with appropriate parameterization, both $t$-SNE and UMAP yield significantly better clustering results than other algorithms.

\section{Providing to a classical data table with a Riemannian manifold structure}
\label{sec:provi}

UMAP method was designed to improve the main limitations of the $t$-SNE method.  $t$-SNE means $t$-distributed Stochastic Neighbor Embedding and it was proposed by Laurens van der Maaten, see all the detail of this method in \cite{maaten}. UMAP algorithm is competitive with $t$-SNE for visualization quality and it improves $t$-SNE limitations. UMAP (Uniform Manifold Aproximation and Projection) is an algorithm for dimension reduction based on algebraic topology, topological data analysis and Riemannian geometry. It was proposed by the Mathematician Leland McInnes in \cite{mcinnes}. UMAP works in a similar way to $t$-SNE, it finds distances in a space with many variables and then tries to reproduce these distances in a low-dimensional space. But UMAP does it very differently because more than distances it tries to reproduce the topology, not necessarily the geometry. UMAP assumes that data is distributed along a Riemannian manifold. A manifold is a uniform $n$-dimensional geometric shape in which, for each point of this manifold, there is a neighborhood around that point that looks like a flat two-dimensional plane. Riemannian manifolds admit local notions of distances, area and angles. To explain the UMAP method we need to define the notion of $k$-simplex and simplicial complexes.  

Let $\{x_0, \ldots, x_k\}$ be points in $\mathbb{R}^n$. We will assume that these points satisfy the condition that the set of vectors in $\mathbb{R}^n$ represented by the differences with respect to $x_0$, that is $\{x_1 - x_0, x_2 - x_0, \ldots, x_k - x_0\}$
are linearly independent. 

\begin{definition}
The $k$-simplex generated by the points $\{x_0, \ldots, x_k\}$ is the set of all points
$ z = \sum_{i=0}^{k} a_i x_i,$
where
$ \sum_{i=0}^{k} a_i = 1. $
For a given $z$, we refer to $a_i$ as the $i$-th barycentric coordinate.
\end{definition}

Simplicial complexes are generalizations of graphs. A simplicial complex $S$ in $\mathbb{R}^n$ is a set of simplices such that every face of a simplex in $S$ is also a simplex in $S$. The intersection of two simplices in $S$ is a face of each of them.
Given data set presented as a finite metric space, we need to produce a simplicial complex such that the algebraic invariants of the simplicial complex reflect the shape of the data. To do that, we need to make the connection between clustering and components precise, via single-linkage clustering, which works as follows.

\begin{figure}[h]
\sidecaption
\includegraphics[scale=.30]{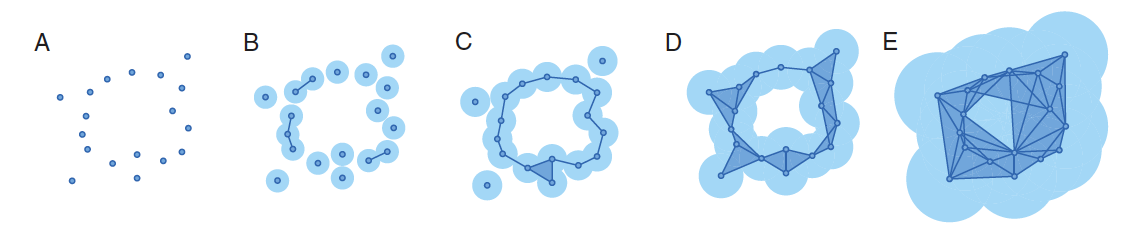}
\caption{As \(\epsilon\) increases, more and more simplices are added to the simplicial complex and topological features emerge. In panels \(C\) and \(D\), a circle can be detected.}
\label{fig:A}       
\end{figure}

\begin{enumerate}
    \item Choose a parameter \(\epsilon\).
    \item Assign two points \(x\) and \(y\) to the same group if they are connected by a path of points (for some \(k\))
    $
    x = x_{0}, x_{1}, x_{2}, \ldots, x_{k-1}, x_{k} = y
    $
    such that each point \(x_{i}\) is at a distance \(\epsilon\) from \(x_{i+1}\). See the Figure \ref{fig:A}.
\end{enumerate}

The Nerve Theorem and its corollary \cite{oudot}  are the fundamental theoretical basis that allows to go from topological spaces to simplicial complexes and then to data. The Čech complex allows to demonstrate that there exists a homeomorphism between the union of balls (determined by the parameter \(\epsilon\)) and the nerve and therefore we will have a bijection between the data and the simplicial complexes.

\begin{definition}
Let \(X \subset \mathbb{R}^{n}\) be a finite subspace and fix \(\epsilon>0\).
The Čech complex \(C_{\epsilon}\left(X, \partial_{X}\right)\) is the simplicial complex with vertices the points of \(X\), and
a \(k\)-simplex \(\left[v_{0}, v_{1}, \ldots, v_{k}\right]\) when a set of points   \(\left\{v_{0}, v_{1}, \ldots, v_{k}\right\} \subset X\) satisfies
$
\bigcap_{i} B_{\epsilon}\left(v_{i}\right) \neq \emptyset.
$
\end{definition}

\begin{theorem} [Nerve Theorem corollary] 
Let \(X \subset \mathbb{R}^{n}\) be a finite subspace and fix \(\epsilon>0\). There exists a homeomorphism:
$
\bigcup_{x \in X} B_{\epsilon}(x) \cong\left|C_{\epsilon}\left(X, \partial_{X}\right)\right|
$
between the union of balls and the nerve $N(\mathcal{U})$ (the geometric realization) of the Čech complex.
\label{TN}
\end{theorem}

The above theorem guarantees that there exists a homeomorphism between the union of balls and the nerve, so, there is relation one-to-one (bijection) between data and Čech complex. To apply these ideas, UMAP choose a radius from each point, connecting points when those radii overlap, then we can create a simplicial complex using 0, 1, and 2 simplexes as points, lines, and triangles. Choosing this radius is critical, too small choice will lead to small, isolated clusters, while too large choice will connect everything together. UMAP overcomes this challenge by choosing a radius locally, based on the local distance to each point to the $k$-th nearest neighbor. To do that, Riemannian metric is used, see the definition in \cite{rabadan}. The choice of $k$ determines how locally we wish to estimate the Riemannian metric. A small choice of $k$ means we want a very local interpretation, while, choosing a large $k$ means our estimates will be based on larger regions. {\em This is very important, because it means that the UMAP algorithm provides the data table with local distance notions}. To do that, UMAP minimized the Cross Entropy, see \cite{mcinnes}.

\section{Riemannian statistics for any type of data}
\label{sec:riem}

Defining statistical methods on Riemannian manifolds poses a unique challenge due to the absence of fundamental vector space operations like addition and scalar product. A critical error would arise from employing statistical indices grounded in the Euclidean space structure of \( \mathbb{R}^n \). To illustrate, consider the scenario where one intends to furnish the UMAP method with a correlation circle. To illustrate, we will utilize the following data table \ref{tab:2}, which includes the school grades of ten students.

\begin{table}
\caption{Students data}
\label{tab:2} 
\begin{tabular}{p{2cm}p{1.8cm}p{1.8cm}p{1.8cm}p{1.8cm}p{1.8cm}}
\hline\noalign{\smallskip} 
& Math & Science & Spanish & History & Phys. Ed. \\ 
\noalign{\smallskip}\svhline\noalign{\smallskip}
Lucía & 7.0 & 6.5 & 9.2 & 8.6 & 8.0 \\ 
Pedro & 7.5 & 9.4 & 7.3 & 7.0 & 7.0 \\  
Inés & 7.6 & 9.2 & 8.0 & 8.0 & 7.5 \\  
Luis & 5.0 & 6.5 & 6.5 & 7.0 & 9.0 \\
Andrés & 6.0 & 6.0 & 7.8 & 8.9 & 7.3 \\  
Ana & 7.8 & 9.6 & 7.7 & 8.0 & 6.5 \\  
Carlos & 6.3 & 6.4 & 8.2 & 9.0 & 7.2 \\ 
José & 7.9 & 9.7 & 7.5 & 8.0 & 6.0 \\  
Sonia & 6.0 & 6.0 & 6.5 & 5.5 & 8.7 \\ 
María & 6.8 & 7.2 & 8.7 & 9.0 & 7.0 \\ 
\noalign{\smallskip}\hline\noalign{\smallskip}
\end{tabular}
\end{table}

\begin{figure}[ht]
\sidecaption
\includegraphics[scale=.18]{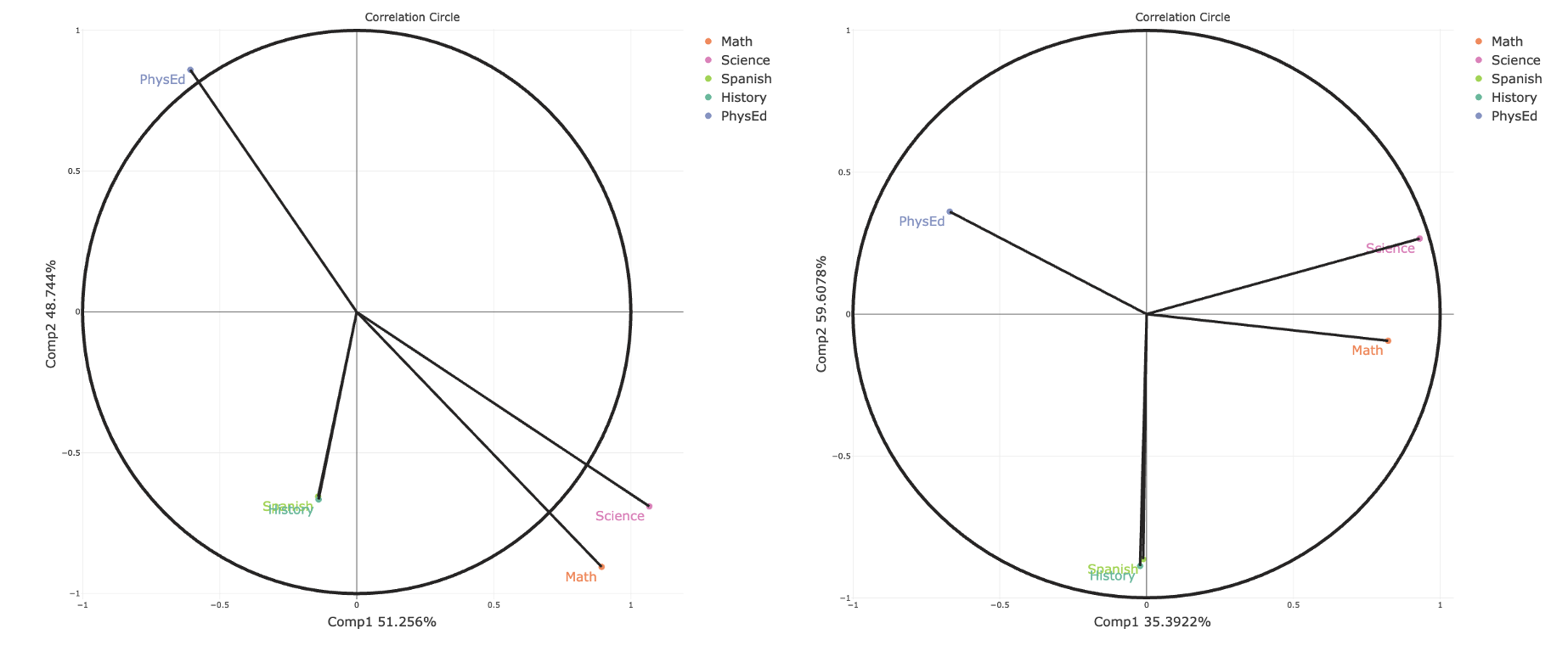}
\caption{UMAP circle of correlation with Pearson correlation on left and with Riemannian correlation on right.}
\label{fig:B}       
\end{figure}

In Principal Component Analysis the coordinate in the correlation circle of variable \(X^{j}\) on axis \(r\) is given by
$
R\left(X^{j}, C^{r}\right)
$
which is the correlation coefficient between the \(j\)-th variable and the \(r\)-th principal component. Using this idea, if we plot UMAP correlation circle using Pearson correlation index, the result is shown in left panel of Figure \ref{fig:B}. Clearly, the correlation circle on the left in Figure \ref{fig:B} exhibits a significant error, namely, certain variable arrows extend beyond the sphere of radius 1. That is to say, we don't have the property:
$
R^{2}\left(X^{j}, C^{s}\right)+R^{2}\left(X^{j}, C^{r}\right) \leq 1.
$
This discrepancy arises from computing correlations as if the data were in a Euclidean space, employing the classical index of correlation. However, the data resides on a Riemannian manifold, with local distances generated by UMAP. Consequently, there is a necessity to define Riemannian correlation, Riemannian mean, and, more broadly, necessitating the development of {\em Riemannian Statistics}. In the subsequent definition, we generalize Fréchet's mean:

\begin{definition}
Let $X \in M_{n \times p}$ the data table. We denote by $\vec{x}_{1}, \ldots, \vec{x}_{n} \in \mathbb{R}^p$ the rows of $X$ and by $\vec{y}_{1}, \ldots, \vec{y}_{p} \in \mathbb{R}^n$ the columns of $X$. Each vector $\vec{x}_{i}$ can be also considered a point in the Riemannian manifold $M$ induced by the simplicial complex. Each pair of vectors $\vec{x}_{i}$ and $\vec{x}_{j}$ has associated a local distances $d_{\textsf{\tiny{UMAP}}}({\vec{x}_i},{\vec{x}_j})$ generated by the UMAP algorithm with its $k$ nearest neighbors \footnote{If the vectors are not in the same $k$ nearest neighbors then a merge distance is used.}. The {\it Riemannian mean} is the minimizer of the sum-of-squared distances to the data $
\vec{g}=\arg \min_{\vec{x} \in M} \sum_{i=1}^{n} d_{\textsf{\tiny{UMAP}}} \left(\vec{x}, \vec{x}_{i}\right)^{2}.
$
\end{definition}

By leveraging the one-to-one relationship given by the Nerve Theorem corollary in \ref{TN} we define the the Riemannian correlation as follows.

\begin{definition}
Let $\vec{x}_{\alpha}$ and $\vec{x}_{\beta}$ rows of $X$, we define the subtraction induced by the UMAP algorithm as 
$\vec{x}_{\alpha} \ominus  \vec{x}_{\beta}=\rho_{\alpha\beta}(\vec{x}_{\alpha} - \vec{x}_{\beta}),$ where $\rho_{\alpha\beta}$ is computed as follows\footnote{Note that $\rho_{\alpha\beta}$ is not a parameter of the model, it can be calculated thanks to the one-to-one relationship given by the Nerve Theorem in \ref{TN}.}

$$
\rho_{\alpha\beta} = \left\{\begin{array}{ccc} 
\frac{d_{\textsf{\tiny{UMAP}}}(\vec{x}_{\alpha},\vec{x}_{\beta})}{d(\vec{x}_{\alpha},\vec{x}_{\beta})} & \textbf{ if }  & d(\vec{x}_{\alpha},\vec{x}_{\beta}) \neq 0 \\ \\
1 & \textbf{ if } & d(\vec{x}_{\alpha},\vec{x}_{\beta}) = 0
\end{array}\right.
$$

\noindent with $d$ the Euclidean distance in $\mathbb{R}^p$. We defined the variance-covariance matrix $ S \in M_{p \times p}$ of $X$ as
$ S = \frac{1}{n} \sum_{i=1}^n (\vec{x}_{i} \ominus \vec{g}) (\vec{x}_{i} \ominus \vec{g})^t,$ where $(\vec{x}_{i} \ominus \vec{g}) = \rho_{i\lambda} \left(x_{i 1}-\vec{g}_{1}, \ldots, x_{i p}-\vec{g}_{p}\right)^t$.

\noindent so, we define \textit{Riemannian correlation} between $y_i$ and $y_j$ columns of $X$, that are in $\mathbb{R}^n$, as follows
$R(\vec{y}_i,\vec{y}_j)=\frac{S_{ij}}{\sqrt{S_{ii}S_{jj}}}$.
\end{definition}

Now we have the property \(R^{2}\left(X^{j}, C^{s}\right)+R^{2}\left(X^{j}, C^{r}\right) \leq 1\) and therefore if we plot again the UMAP  correlation circle using the  Riemannian correlation index, the result shown on right panel on Figure \ref{fig:B} is now correct. 

\section{Conclusions and Future Work}
\label{sec:conc}

In this paper, we successfully extend the ideas proposed by \textit{Pennec et al.} in \cite{pennec}, broadening the scope to compute Riemannian statistical indices and Riemannian data analysis models to any data table. Unlike previous approaches, our methodology is not restricted to data with an intrinsic Riemannian manifold structure. This advancement opens up a new field of research, where diverse methods like regression, $k$-means, and more, can be generalized for broader applicability.

Currently, we are actively engaged in implementing these novel ideas in both \textbf{R} and \textbf{Python}, ensuring practical adoption and seamless integration across different computational platforms.

\input{referenc}

\end{document}

%% file: referenc.tex
%
%
%